\documentclass[english]{article}
\usepackage{geometry}
\usepackage[T1]{fontenc}
\usepackage[latin1]{inputenc}
\usepackage{graphicx}
\makeatletter
\usepackage{babel}
\makeatother

\begin{document}

\title{Superradiance and stimulated scattering in SNR 1987A}
\author{Jacques. Moret-Bailly
\footnote{Physique, Universit\'e de Bourgogne, Dijon France. email: jacques.moret-bailly@u-bourgogne.fr}}	

\maketitle
 \begin{abstract}
   The rings observed around supernova remnant 1987A are emitted by a plasma mainly made of ionized and neutral hydrogen atoms. With a density of 10 power 10 atoms per cubic metre, and at least a dimension of plasma of 0.01 light-year, the column density is 10 power 24 atoms per square metre, much more than needed for an optically thick gas at Lyman frequencies (Case B). While, at 10000 K, the bulky gas would absorb Lyman lines fully, at 50000K it emits superradiant lines. As superradiance de-excites the atoms strongly, nearly all available energy is emitted in a few competing modes: Superradiance appears only for beams which cross the largest column densities; for an observation from Earth, these beams generate three elliptical hollow cylinders whose bases are the observed rings; admitting that the Earth is not in a privileged direction, these cylinders envelope ellipsoidal shells observed, for the external rings, by photon echoes. For the equatorial ring, the brightness of the superradiant beams is multiplied by a quasi-resonant induced scattering of the rays emitted by the star. The de-excitation of atoms by emitted beams cools the gas strongly, so that ionization decreases fast, the process self accelerates. The energy of the high radiance rays from the star is strongly scattered to the ring while the low radiance of the glow which surrounds the star is rather amplified. The fast absorption of the radial beams produces the radial decrease of radiance of the ring while a competition of modes produces the pearls.
 \end{abstract}
keywords: Radiation mechanisms: non thermal; Radiative transfer; Supernovae: remnants, individual.

\section{Introduction}
Various codes have been set to study interaction of light with interstellar matter (Ferland et al. 1998, R\"ollig et al. 2007); but the number of parameters is so large that only particular cases were studied; for instance Ferguson et al. 1996 wrote \textquotedblleft for this approximation, all induced processes [...] are ignored\textquotedblright{} . In \textquotedblleft Case B\textquotedblright{} of Baker \& Menzel (1938), neutral hydrogen column density is large enough for Lyman lines to be optically thick, but this case is mainly studied at relatively low temperatures, typically 10$^4$ K (Hummer \& Storey 1992).

At higher temperatures, and high light intensities, Case B is mainly an interaction with a bulky matter similar to a laser medium without population inversion. The aim of this paper is taking into account superradiance and coherence of light-matter interactions in the region which emits the rings (necklaces) of supernova remnant 1987A.
 
 \medskip
The magnitude of Supernova 1987A decreased from its maximum 2,9 to 16 in 1000 days (Arnett et al. 1989). The initial spectrum was very rich in UV and showed broad hydrogen and helium emission lines; then UV light decreased and it remains only a diffuse emission of lines of low ionization elements (Woosley et al. 1988).

Ground based images of SN1987A showed a weak blob of gas which was resolved as a circumstellar \textquotedblleft  equatorial ring\textquotedblright{}   (ER) by the ESA Faint Object Camera on Hubble on August 1990. Comparing time delay between maximal emissions of supernova and ring gave a diameter of the ring: 1.2 light-year (ly); knowing its angular size, the distance of the supernova was set (Panagia et al. 1991).

In 1994, two larger \textquotedblleft outer rings'' (OR) were detected; in 1997, spots (pearls) appeared on the ER, then some much brighter, \textquotedblleft hot spots'' appeared at inner rim of the ring. Except for its small irregularities, the pearl necklace is very similar to figures of mode selection in laser systems emitting light on cones, while the fast time and space variation of radiance of the hot spots suggests a non-linearity rather than complex explanations criticized for instance by Lloyd et al. (1995); the source, made of excited, neutral hydrogen atoms may be generated by shock waves or by photoionization, like the spectra of planetary nebulae (Plait et al. 1995).

\medskip
 Inside the rings, we mainly follow the photoionisation model developed by Chevalier et al. (1995), Lundqvist et al. (1996) and Lundqvist (1999), but our model differs from their model because we suppose that the star remains very hot and bright, emitting in each direction a luminous flux equivalent to the flux received from the necklace.

\medskip
Section \ref{in} recalls the emission and diffusion of light inside the ER.

Section \ref{ER1} of pure optics, recalls stimulated emission of light, leading to superradiance and saturation.

Applying the results described in section \ref{ER1}, section \ref{ER2} shows how the radiance produced by superradiance is increased and extended to a continuous spectrum by an induced, coherent, nearly resonant scattering of light emitted by the star, scattering which absorbs almost completely light of the star, but not the glow which surrounds it. 

Section \ref{ER3} explains the variations of radiance in the equatorial ring, in particular it explains how \textquotedblleft hot spots\textquotedblright{}   appear.

Section \ref{Geo} studies the geometry of the rings.

{\it All ellipsoids used in this paper are invariant by rotations around an axis, axis z' when it is defined.}

\section{The plasma inside the equatorial ring (ER).}\label{in}

After the blue supernova explosion, the remaining star had the spectrum of a blue star, emitting very broad Lyman lines of H$_{I}$ and He, and a strong continuous spectrum of shorter frequencies (extreme UV). We suppose that, after the main supernova glow, the surface temperature of the star did not decrease very strongly, so that its spectrum remains mainly made of less wide Lyman lines of H$_{I}$ and extreme UV. 

To simplify the explanations, consider, at first, only the main ring (ER), and, in a first approximation, suppose that the system has a spherical symmetry. The cooling of the plasma is limited by the absorption of extreme UV which ionizes H$_{I}$ into electrons and protons (H$_{II}$), building a spherical bubble of H$_{II}$ inside a cloud of H$_{I}$ (Chevalier et al. 1995). In the bubble, various ionized atoms (He, C, N, O, ...) absorb energy, radiate lines and produce a Rayleigh scattering of Lyman emission lines. The largest fraction of this scattering is coherent, producing in that way refraction; weaker incoherent scattering and emissions produce a glow around the star, similar to the blue of our sky. As protons and electrons are nearly free, they do not play a notable optical role. 

\begin{figure}
\includegraphics[height=6 cm]{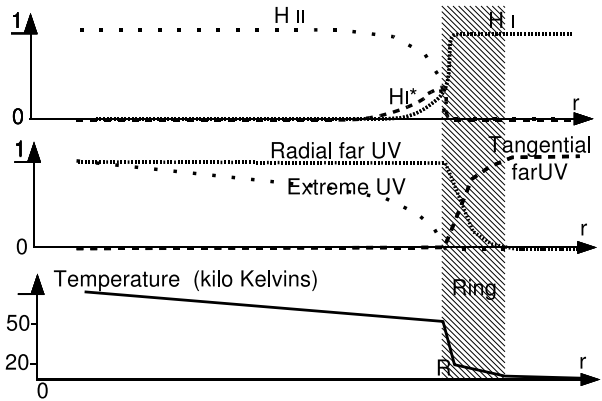}
\caption{\label{rad} Variation of the relative densities of H$_I$, H$_{II}$ and excited atomic hydrogen H$_I$*, relative intensities of light, and temperature along a radius starting at the star.}
\end{figure}

For a chosen orientation, locate a point by its distance r to the center of the star, and set r = R at the inner point of the necklace (figure \ref{rad}). For r larger than about 3R/4, the remaining intensity in extreme UV has decreased enough, and the ionization by several steps is improbable enough, to let appear some mainly excited  H$_{I}$, with a density increasing with r, so that Lyman lines are scattered and emitted more and more strongly (Sonneborn et al. 1998). 

\section{Reminder on stimulated emissions of light.}\label{ER1}

Planck's formula for the electromagnetic energy of a monochromatic mode inside a blackbody connects the spectral radiance (or the monochromatic density of energy) in a light mode with temperature of the blackbody which becomes, by definition, temperature of the mode. Spectral temperature of a transition between two states a and b, of energies $w_a > w_b$ in a medium where the respective densities are $N_a$ and $N_b$ defines a temperature of this set of states by the relation $N_a/N_b = exp((w_b-w_a)/kT)$, if there is not an inversion of populations. All optical interactions, in particular stimulated emissions and absorptions, tend to increase the entropy of a system by an equalization of temperature of light with temperature of involved  molecular transition. 

Einstein \cite{Einstein} introduced stimulated emission of light; studies on lasers show that they start by a spontaneous emission which may be considered as stimulated by the zero point field. In thermodynamics, emission and absorption are two faces of an unique interaction which tends to increase entropy: if a gas is hotter than the surrounding medium, it radiates light, else, it absorbs. Here, we mainly consider the first case and, in a first approximation, strong Lyman transitions. 

Depending on the column density of emitting molecules, three types of emission may be distinguished, and easily observed in a hot gas:

- for a low column density, the field remains nearly the zero point field, so that its increase of intensity (this increase is the usual intensity) is proportional to the path of light. This \textquotedblleft Case A\textquotedblright{} is the most common in the labs.

- for a higher column density, assuming a negligible radiative de-excitation of the excited levels, the intensity becomes an exponential function of the path. The final intensity depends much on the amplification coefficient. For instance, for a mainly Doppler-broadened line, the density of molecules having a radial speed in a fixed small interval $\delta V$ is maximal around radial speed $V=0$, so that the amplification coefficient is maximal at the center of the line which, therefore, is sharpened. It is  usual superradiance.

- for a high column density, in \textquotedblleft Case B\textquotedblright{} , the depopulation of the excited state of the studied transition:
 
i) limits the intensity close to the center of the line; 

ii) decreases the spontaneous or induced emission of the line in other modes; it is a  \textquotedblleft  competition of modes'' which limits the number of high intensity emitted light beams. Using mirrors to increase virtually some paths, and therefore select monochromatic modes, a laser illustrates this behavior.

\section{Generation of the ER by superradiance.}\label{ER2}
As previously described incoherent scatterings are weak, far UV light propagates in ionized hydrogen without a large absorption until it reaches a region cold enough to decrease strongly ionization of hydrogen at low pressure; around 50 000 K hydrogen plasma contains enough neutral atoms, mainly in excited states, to radiate Lyman lines strongly; this temperature was observed by Arnett et al. (1989) close to the necklace.

The measured density of hydrogen ions and atoms around the necklace is of the order of 10$^{10}$ m$^{-3}$, which gives, for paths of 0.01 light-year, that is 10$^{14}$ m, a column density of 10$^{24}$ m$^{-2}$. Compare this amplifying medium with the plasma of a gas laser medium: In an Helium-Neon or ionized Argon laser the number density of atoms is of the order of 10$^{22}$ m$^{-3}$, and the length of active medium, multiplied by the gain of the cavity, of the order of 100 m, so that the number of atoms per unit of surface perpendicular to the light beam (column density) is of the order of 10$^{24}$ m$^{-2}$. 

Temperature decreasing with the distance to the star, more and more protons combine with electrons to increase the density of hydrogen atoms mainly in excited states . Oscillator strength of H$_ I$ is large for Lyman lines (0.8324 for Ly $\alpha$), so that column density of excited atoms reaches a value for which the medium becomes able to amplify Lyman frequencies much, generating a strong superradiant emission.

Light reaching the Earth was emitted in a cone, almost a cylinder whose base is the light-emitting ring. The height of the emitting region of this cylinder must be larger than its thickness to obtain superradiance in direction of the Earth. Assuming that the Earth is not privileged for the observation of SNR 1987A, all emitting regions merge into an envelope of the cylinders, which is the shell found in the ionization theory, compatible with the shock wave theory. 

By a selection of modes, it remains only, in a given direction, a limited number of beams emitted nearly tangentially to the inner surface of the shell. The geometry of a bright elementary mode emitted to a telescope, whose Clausius invariant is $n^2\lambda^2$, is defined by the mirror of the telescope and a  region of the ring exactly resolved by the telescope; the similarity with the conical laser emissions appears;  however, inhomogeneities of the stellar medium introduce larger perturbations. 

\section{Amplification of the emission of the ER by induced scattering.}\label{ER3}
Superradiance cannot explain an emission of very broad lines or continuous spectra. Incoherent, quasi-resonant scattering can, but it is weak. Induced by superradiant beams, scattering becomes strong in privileged directions, much stronger, than an absorption followed by an induced emission at line frequencies. The scatterings induced by feet of superradiant lines broadens them more and more. A combination of this broadening and saturation of scattering finally transfers nearly all light at all wavelengths from radial light emitted by the star, to tangential light, building necklace and pearls. 

The scattering tends to equilibrate temperatures (therefore radiances) of involved beams; the solid angles of observation of a pearl necklace is much larger than the solid angle of observation of the star; thus, the flux received from the star is so smaller that it is invisible.

Around the star, the halo of luminescence of various ionized atoms is observed through large enough a solid angle; as its radiance is low, it is not absorbed by the shell, maybe amplified by it; it remains visible while the star disappears.

Superradiance depopulates the excited states of atoms H$_{I}$  which, colliding protons, slows them down, cooling hydrogen strongly and allows capture of an electron; generated atoms are   quickly de-excited by superradiance. This reaction process may amplify superradiance almost linearly, or reach the catastrophic point at which almost all H$_{II}$ is locally transformed into H$_{I}$; in this last case, the radiance of tangential emission bursts, a hot spot appears at the inner rim of the ring. In both cases, radiance decreases from the inner rim because the intensity of radial radiation decreases by scattering or absorption (Plait, et al. 1995, Panagia et al. 1991).

\medskip
\begin{figure}
\includegraphics[height=5 cm]{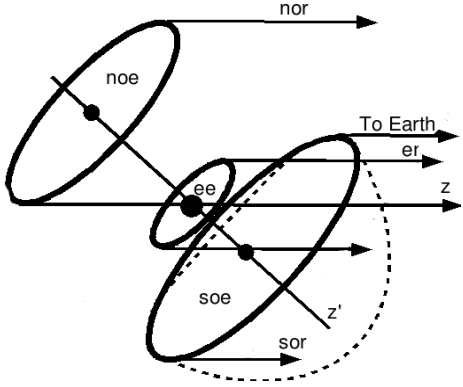}
\caption{\label{ellip} Elliptic sections, by the plane containing axis of symmetry z' and direction z to Earth, of the ellipsoidal shells of atomic hydrogen which generate the necklaces (the ellipsoids are generated rotating the ellipsis around z' axis); noe, ee and soe: north outer, equatorial and south outer ellipsoids; nor, er, sor: traces of the hollow, tangent cylinders of light oriented to Earth; the dotted lines suggest how, taking into account the variations of density of the gas, the outer shells can get a more realistic shape of pear (they should be resized).}
\end{figure}

The tangential rays directed to the Earth draw the circular ring and competition of modes selects some of them, stabilized by slowly varying perturbations of the system, for instance by dust: the selected modes make the \textquotedblleft  pearls''. Remark that mathematical definition of the modes does not imply that they are monochromatic: a mode is a ray in the space of solutions of a linear set of equations (here Maxwell's equations). This is happy because a perfectly monochromatic mode never starts and stops, it is not physics! Here, the nonlinearity of the superradiant process provides an interaction between modes which could be defined for different frequency bands.

The columns of light generated by the superradiant modes excite various atoms, generating columns of excited atoms. These cold atoms radiate nearly monochromatic, superradiant modes collinear with the initial UV modes, increasing the spectral complexity of the rings.

\section{Geometry of the rings.}\label{Geo}
\begin{figure}
\includegraphics[height=6 cm]{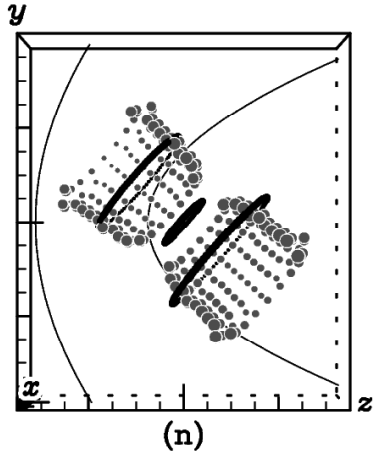}
\caption{\label{echo}Section of the \textquotedblleft  circumstellar hourglass'' scattering shells found from echoes by Sugerman et al. and the rings (fig. 43n of Sugerman  et al. 2005). Compare with fig. \ref{ellip} obtained from our hypothesis.}
\end{figure}
The shell which generates equatorial ring os not spherical, but ellipsoidal, oblate (figure \ref{ellip}); as the observed density of gas along its short axis z' is larger than in other directions, this shape is easily explained, for instance in the photoionization model: assuming that radiation of the star is isotropic, column density of ionized hydrogen must be equal in all directions, so that the path to the shell is smaller along z'. 

Such shells were found and sketched for the external rings by Sugerman et al. (2005) who, to represent the emitting zones, darkened regions where the shells cross tangentially the cylinders of light oriented to the Earth; their figure 43n, reproduced as figure \ref{echo} shows a metal cup shape close, in precisely observed regions, to ellipsoids; a common axis of symmetry z' for ejected gas and centers of the external shells suggests a common origin; stars at these centers are needed to explain easily a generation of the shells. These stars may be neutron stars invisible because, heated by accretion of gas, but small, they radiate mainly UV. Halton Arp (1999) writes that alignments of stars are not all due to accidental observations, resulting probably from explosions, here the first explosion of the fast spinning core of the supernova.

Around the neutron stars, the variation of gas density in z' direction is asymmetric, so that the shell takes the shape of a pear, the external necklaces loosing their centers of symmetry.

\section{Conclusion.}
Taking into account coherent interactions of light with the gas which surrounds SNR 1987A solves the main problems:

- Coherent optics and thermodynamics in a simple shell of  H$_I$ explain the observation of a bright ring with pearls, its radial decrease of brightness and hot spots on its internal rim.

- Thermodynamics of absorption and scattering in the shell of H$_I$ explains the progressive disappearing of the central star while its glow generated by low ionization ions remains.

The three emitting shells may be generated by three aligned stars similar to the systems observed by Arp (1999).

\medskip
Except for this last explanation, our results are directly deduced from observations, without any new hypothesis. It does not require any new physics either. It appears to us that all observations have simple explanations. Therefore the coherent light-matter interactions which should be taken into account in astrophysics should not be limited to refraction.

Does generation of a ring by photoionization, superradiance and stimulated scattering applies to some other observed rings, rather than gravitational lensings which require improbable alignments of stars with the Earth?

\label{lastpage}
\end{document}